# Single-beam reflection technique for determination of nonlinear-refractive index of thin-film semiconductors using an electrically focus tunable lens

**Juan Serna[1,*], Julián Henao[2], Edgar Rueda[2], Abdullatif Hamad[3], and Hernando Garcia[3,*]**

[1] Grupo de Óptica y Espectroscopía, Centro de Ciencia Básica, Universidad Pontificia Bolivariana, Cq. 1 No 70-01, Campus Laureles, Medellín, Colombia
[2] Grupo de Óptica y Fotónica, Instituto de Física, Universidad de Antioquia, Calle 70 No. 52-21, Medellín, Colombia.
[3] Department of Physic, Southern Illinois University, Edwardsville, Illinois, USA.

Email: juan.sernar@upb.edu.co

**Abstract.** In this paper, we propose a technique named reflection F-scan or RF-scan, that can be used to measure the nonlinear-refractive index $n_2$ of thin-film semiconductors. In this technique, a *p*-polarized Gaussian beam is focused using an electrically focus-tunable lens onto a sample, which is positioned at a fixed distance from the lens and makes an angle with respect to the optical axis. Due to EFTL has the capability to vary its focal distance over a specific range when an electric current is applied to it. The electrically focus-tunable lens varies its focal distance as a function of an applied electric current over a specific range thus, when light is focused on the surface of the sample the beam intensity is high enough to generate nonlinear optical effects such as changes in the refractive index of the material. This changes are then register as variations in reflectance, measured by an intensity detector. Results for three-dimensional $CH_3NH_3PbBr_3$ hybrid perovskite thin films are presented.

## 1. Introduction

Since the advent of the laser there has been a growing interest on the determination of the nonlinear-optical properties of known and new materials, due to its technological possibilities in a wide range of applications that include for example optical switching, optical processing, optical modulation, communications, and medical treatment and diagnosis [1–3]. In particular, to measure the nonlinear absorption and nonlinear-refractive index a technique call Z-scan has been widely used, especially because of its relatively simple optical setup and data treatment, which is based on the measurement of the variations of the intensity of a focus Gaussian beam after it is transmitted through the sample [4]. These technique has to main variations, the open-aperture setup where all the transmitted light is gathered into a photodiode detector to measure the nonlinear absorption, and the closed-aperture setup where an iris in the optical axis is included to detect phase variations produced by the nonlinear-refractive index. Because of the used of an iris, closed-aperture setup is less robust to misalignments. To overcome this difficulty the reflection z-scan (RZ-scan) technique was proposed to measure nonlinear properties of opaque materials or nonlinear properties at the material surface. In this technique, all the light reflected by the sample is collected and measure by the detector, thus any change in reflectance is directly related to a change of the refractive index of the material, at the surface. This setup has the advantage that no iris is needed, making the measurement of the nonlinear-refractive index as simple as the measurement of the nonlinear absorption in a conventional z-scan. The sensibility of the RZ-scan has been also improved by moving the incident-beam angle closed to the Brewster angle [5–8]. More recently, a variation to the Z-scan, call F-scan, has been proposed by introducing electrically focus-tunable lens (EFTL) instead of a translation stage for the sample. The main advantage of this variation is that the sample is fixed in a point, i. e., there are no moving parts, making the system robust to misalignments. Yet, to measure the nonlinear-refractive index with the closed-aperture technique still requires an iris in the optical axis, and a free-space beam propagation is still needed. In this work, to measure the nonlinear-refractive index we have implemented the analog to the RZ-scan technique for the case of the F-scan. In the following, we will explain the main issues in order to calibrate the EFTL and its relation to the beam quality factor $M^2$, we will develop the model analytical expression, and we will experimentally measure the nonlinear-refractive index for a 3-D $CH_3NH_3PbBr_3$ hybrid perovskite thin films..

## 2. EFTL Characterization

For F-scan techniques, such as TF-scan (transmission F-scan) [9], DF-scan (differential F-scan) [10] and RF-scan (reflection F-scan) it is crucial to know how the focal length varies as function of a current applied to EFTL, in order to obtain correct values of two-photon absorption coefficient $\beta$ and nonlinear-refractive index $n_2$. Different techniques can be used to obtain the correct dependence. We used a laser-beam profiler to measure the focal length $f$ as a function of the applied current (circles in Fig. 1), then, using the relation $\wp = 1/f$, the EFTL optical power $\wp$ is obtained as a function of current. In our case, the experimental data was fitted obtaining the following relation (continuous line in Fig. 1):

$$\wp = 0.052 J + 5.3 \tag{1}$$

where $J$ is the applied current measured in mA.

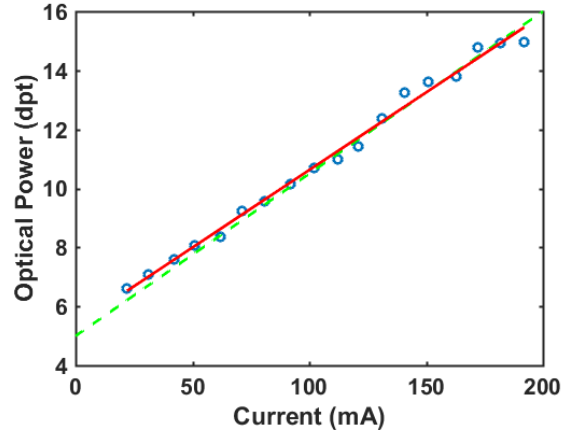

**Figure 2.** The optical power $\wp$ of the EFTL as a function of the applied current. (circles) Experimental data; (continuous line) fit of the experimental data using Eq. (1); (dashed line) data provided by OPTOTUNE Inc.

Another important experimental parameter for the correct determination of the nonlinear-optical parameters is the beam-waist radius $w_0$. Typically, for spherical lenses, and assuming that the beam has a spatial Gaussian profile, the radius of the beam at the beam waist is determined as a function of the focal length using the equation:

$$w_0 = \frac{2\lambda f}{\pi D} \tag{2}$$

where $\lambda$ is the wavelength of the incident beam with spot diameter $D$. More generally, and taking into account aberrations of the Gaussian beam, the beam radius $w(f)$ can be write as [11]:

$$w(f) = \sqrt{\left(M^2 \frac{2d\lambda}{\pi D}\right)^2 + w_i^2 \left(1 + \frac{d}{R_i} - \frac{d}{f}\right)^2} \tag{3}$$

Here, $M^2$ is known as the beam quality factor, $R_i$ is the radius of curvature on the EFTL and $d$ is a fixed distance measured from EFTL. At the beam waist, i.e., when $d = f$, we can approximate Eq. (3) to:

$$w(f) = w_0^{(c)} = M^2 \frac{2f\lambda}{\pi D} \tag{4}$$

Eq. (4) shows that the beam waist is corrected by the beam quality factor $M^2$. This correction factor has into account the existence of optical aberrations on the optical system that distort the wave-front of the beam. Eq. (2) is then replaced by:

$$w_0 = \frac{2\lambda f}{\pi D} M^2 \qquad (5)$$

To determine the correction factor we used a laser beam profiler to measure the beam waist at each focal plane. Then, by using $M^2$ as the fitting parameter between the experimental data and Eq. (5), a correction factor $M^2 = 1.2$ was obtained for the special case of our EFTL. It is worth mentioning that the value of $M^2$ ensures that the laser beam that impinges on the sample can be treated as an ideal Gaussian beam. Figure 2 shows the difference between corrected and non-corrected beam-waist diameter values as a function of the EFTL focal length.

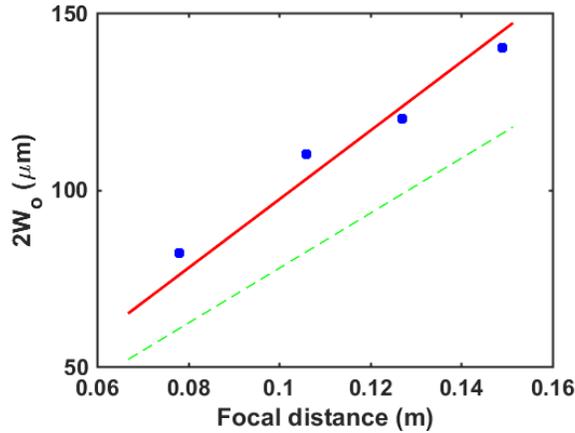

**Figure 2**. Beam-waist diameter as a function of EFTL focal length. (square) Experimental data measured with a laser beam profiler; (dashed line) beam waist diameter calculated with Eq. (2); (continuous line) beam waist diameter calculated with Eq. (5) and $M^2 = 1.2$.

Once the beam-waist radius is correctly determined, it is possible to calculate with precision the beam radius $(f)$ (Eq. (4)) at the sample surface for every programed EFTL focal length,

$$w(f) = w_0 \sqrt{1 + \frac{(d_s - f)^2}{z_0(f)^2}} \qquad (6)$$

where $z_0(f) = \frac{\pi w_0^2(f)}{\lambda}$ is the Rayleigh range. Figure 3 shows the dependence of the beam radius at the sample location as a function of the EFTL focal length. Notice the difference between the results obtained with and without the corrected beam-waist radius. Also notice that for $f$ values smaller than $d_s$ the radius beam at the sample increased faster that those with $f$ values larger that $d_s$. This result is very important when we calculate the intensity at the sample, which is given by expression:

$$I_{in}(r,f) = I_o(f) exp\left(-2\frac{r^2}{w^2(f)}\right) \qquad (7)$$

where $r$ is the radial position transverse to the beam axis, and $I_o(f)$ is given by

$$I_o(f) = \left[\frac{2ln(1+\sqrt{2})}{\tau \cdot \nu \cdot \pi \cdot w_0^2(f)}\right] \qquad (8)$$

In Eq. (8), $\tau$ is the pulsed width of laser, and $\nu$ is the repetition rate.

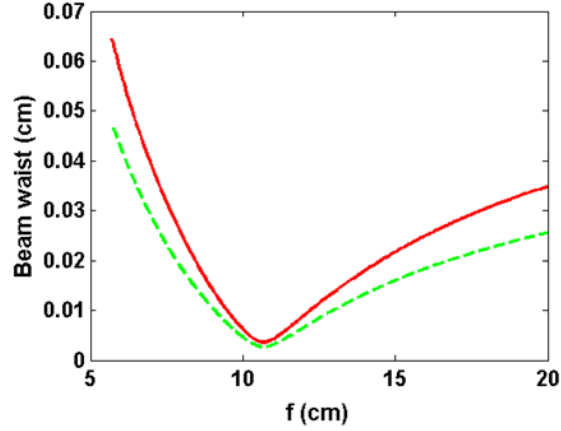

**Figure 3.** Beam radius at the sample as a function of the EFTL focal length. The continuous and dashed lines are the corrected (Eq. (3)) and not corrected (Eq. (2)) beam waist radius $w_0$ respectively. For this plot we used $D = 2.0$ mm and $d_s = 10.7\ cm$.

### 3. Nonlinear-refractive index determination

Reflection F-scan (RF-scan) technique is based on the F-scan principle, where a Gaussian beam that is modulated with a chopper, impinges on an EFTL that has the capability to vary its focal distance $f$ over a specific distance when an electric current is applied to it. The EFTL then focuses the Gaussian beam at different focal points and the sample is place at a fixed position $d_s$ inside the focal range of EFTL. For RF-scan the sample is rotated making an angle with respect to the beam propagation optical axis. The beam reflected by the sample is redirected towards the detector and the reflectance is measured. To improve the sensitivity of the technique, the sample is positioned at Brewster angle condition, where the *p*-polarized component of the optical beam is eliminated, and the reflected beam will have a minimum intensity. Any modification in the refractive index of the sample caused by the incident field will drive the reflected beam out of Brewster condition, and it will produce a change in the intensity. In this way, measuring this modification in the reflected beam, we are able to retrieve information about the nonlinear refractive index $n_2$.

The intensity of the beam when it is reflected by the sample can be calculated by means of the amplitude reflection coefficient $R(\bar{n},\theta)$ on the interface air-sample. This coefficient depends on complex-refractive index $\bar{n}$ and the incident angle $\theta$ and can be expressed as [12]:

$$R(\bar{n},\theta) = \frac{\bar{n}^2 \cos(\theta) - \sqrt{\bar{n}^2 - \sin(\theta)}}{\bar{n}^2 \cos(\theta) + \sqrt{\bar{n}^2 - \sin(\theta)}} \qquad (9)$$

In Eq. (9) the refractive index can be written as $\bar{n} = \bar{n}_0 + \Delta\bar{n}(I)$ with $\bar{n}_0$ being the linear complex-refractive index and $\Delta\bar{n} = (n_2 + i\kappa_2)I$ the change in the complex-refractive index for a non-saturable medium, where $n_2$ and $\kappa_2$ represent the nonlinear-refractive index and nonlinear-extinction coefficient respectively. The normalized reflectance $\mathcal{R}_N$ can be defined as the ratio between reflected-beam power with and without the nonlinear effect:

$$\mathcal{R}_N(z,\theta) = \frac{\int_o^\infty |R(\theta)|^2 I(r,z) r dr}{\int_o^\infty |R_0(\theta)|^2 I(r,z) r dr} \qquad (10)$$

where $z = d_s - f$ and $I(r,z)$ is given by Eq. (7). To solve Eq. (10), we can expand $R(\bar{n},\theta)$ in a Taylor series, i.e. $R(\theta) = \bar{R}_0(\theta) + \Delta\bar{n}\frac{\partial R(\theta)}{\partial \bar{n}}\Big|_{\Delta\bar{n}=0}$ with $\bar{R}_0(\theta)$ being the linear complex-reflection coefficient, and found that the reflectance can be written as

$$\mathcal{R}_N(f,\theta) = 1 + \frac{Re\{\delta(\theta)\}}{1 + \left(\frac{d_s - f}{z_0}\right)^2} \quad (11)$$

The quantity $\delta(\theta) = \frac{\bar{n}_2 I_0}{R_0} \left.\frac{\partial R(\theta)}{\partial \bar{n}}\right|_{\Delta\bar{n}=0}$ is defined as the normalized nonlinear-reflection coefficient, and it is related to the ratio between linear and nonlinear contributions of the reflection's amplitude coefficient, in $r = 0$ and $z = 0$. After some algebraic operations, the final expression for reflectance is:

$$\mathcal{R}_N(f,\theta) = 1 + Re\left\{\left(\frac{2\bar{n}_0^3 \cos\theta - 4\bar{n}_0^2 \sin^2\theta \cos\theta}{\bar{n}_0^4 \cos^2\theta - \bar{n}_0^2 + \sin^2\theta}\right) \cdot \frac{(n_2 + i\kappa_2)I_0}{\sqrt{\bar{n}_0^2 - \sin^2\theta}}\right\} \cdot \frac{1}{1 + \left(\frac{d_s - f}{z_0}\right)^2} \quad (12)$$

Eq. (12) is used to adjust the experimental data, assuming $n_2$ as the fitting parameter. The nonlinear extinction coefficient $\kappa_2$ can be obtain knowing the two-photon absorption coefficient $\beta$ and using the relation $\kappa_2 = \frac{\beta\lambda}{4\pi}$ [13].

### 4. Experimental results

Figure 4 shows the RF-scan experimental set-up. For this setup we used a Q-switch laser with a repetition rate $\nu = 11.9$ kHz, pulse width $\tau = 0.5$ ns, laser emission centered at 1064 nm, and an average laser power around 60 mW. The electrically focus-tunable lens was an OPTOTUNE - 1030 with a tunable focus length between 4 cm to 18 cm, controlled by an OPTOTUNE lens driver that gives a maximum current of 300 mA and a resolution of 0.1 mA, given a focal length resolution of 0.017 mm. The laser Gaussian beam is $p$-polarized and is focused at oblique incidence onto the sample, near to the Brewster's angle condition. We collected the reflected beam from the sample by means of a T-integrating sphere. The output signal from the photodetector is filtered with a Lock-in Amplifier, and processed by a computer.

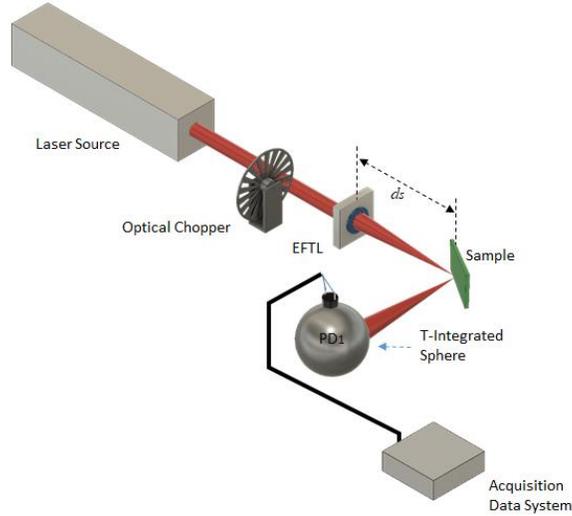

**Figure 4**. Experimental setup for RF-scan. The Gaussian beam is p-polarized by polarizer and focused by the EFTL onto the sample, which is rotated close to Brewster Angle (BA). The reflected light is redirected and collected by a T-integrating sphere, and the signal is process by means of an acquisition data system

To verify the proposed technique, we measured the nonlinear refractive index of a 3-D $CH_3NH_3PbBr_3$ hybrid perovskite thin film, that was prepared according to a previous reported method [14]. We measured the nonlinear extinction coefficient $\kappa_2$ using TF-scan. In this case, the experimental setup was modified in order to collect the transmitted light, as shown in Fig. 5.

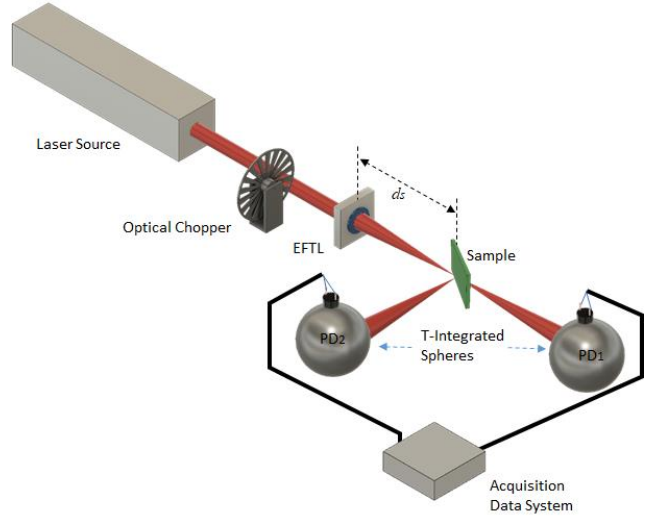

**Figure 5**. Experimental set-up to measure the nonlinear refractive index (RF-scan) and two-photon absorption coefficient (TF-scan).

To calculate the value of $\beta$ we fitted the $T(f)$ transmission experimental data, shown in figure 6(a), using the following equation:

$$T(f) = \frac{1}{B(f)} \int_0^\infty \ln[1 + B(f)\text{sech}^2(\rho)]d\rho \qquad (13)$$

Where $B(f) = \beta(1-R)I_o(f)L_{eff}$. Here $R$ is the reflection coefficient, $I_o$ is the peak intensity of the beam as function of the focal length, $L_{eff} = (1 - e^{-\alpha L})/\alpha$ is the effective sample thickness, $L$ is the sample thickness and $\alpha = 10^{-4}\text{cm}^{-1}$ is the linear absorption coefficient. Finally, $\rho$ is an integration variable expressed as $\rho = 2\ln(1 + \sqrt{2})/\tau$, and $\tau$ is the full width at half-maximum pulse duration. We found that 3-D perovskite presents a $\beta = (-6.0 \pm 2.0) \times 10^{-8} \frac{m}{W}$ and the $\kappa_2 = (-5.18 \pm 1.7) \times 10^{-15} \frac{m^2}{W}$. Using the value for $\kappa_2$, we fitted the experimental data obtained with RF-scan using Eq. (12) as it is shown in Fig. 6(b), obtaining a $n_2 = (-2.0 \pm 0.6) \times 10^{-14} \frac{m^2}{W}$. Some interesting aspects can be observed in the results shown in Fig. 6(b). One is related with the reduction of the reflectivity when the focal point of EFTL is close to the sample: the refractive index suffers a negative change that can be associated to self-defocusing phenomena [15]. The second is that in RF-scan, unlike TF-scan, the real part of nonlinear refractive index produces an amplitude change in the reflected beam [16], this means that it is not necessary the use of spatial filters to determine the value of $n_2$, simplifying the experimental set-up. Finally, using a Brewster angle condition for the incident laser beam, allows to increase the sensitivity of the RF-scan system. In a Brewster angle configuration the *p*-polarized component of electric field is annulated thus, if the refractive index changes, this component of polarization is recovered and the signal detected increases and can be directly related to the existence of a nonlinear refractive index.

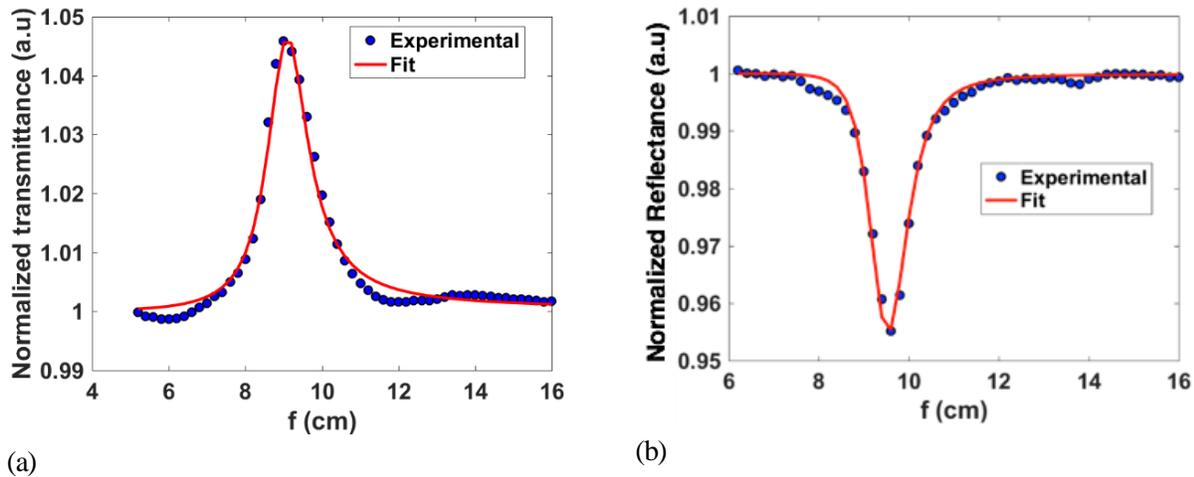

**Figure 6**. Results for 3-D CH$_3$NH$_3$PbBr$_3$ hybrid perovskite thin films for a) TF-scan experiment and b) RF-scan experiment

In conclusion, we have implemented a new technique, RF-scan, to measure the nonlinear refractive-index in thin film samples. This technique, that it is a variation of reflection Z-scan technique, uses an electrically focus-tunable lens to generate a variable focal length. To verify the technique we measured the nonlinear-refractive index for 3-D CH$_3$NH$_3$PbBr$_3$ hybrid perovskite thin film, obtaining $n_2 = (-2.0 \pm 0.6) \times 10^{-14} \frac{m^2}{W}$, which is a typical value for this class of semiconductor materials.